\documentclass{article}
\usepackage{graphicx}
\usepackage{orcidlink}
\usepackage{authblk}
\usepackage{xurl} 
\usepackage{hyperref} 

\title{Changing topic bias in biomedical science maps by linking documents through alternative data sources: policy documents, patents, authors, Facebook, and Twitter
}

\author[1]{Juan Pablo Bascur\orcidlink{0000–0002–4077–1024}}
\author[1]{Rodrigo Costas\orcidlink{0000–0001–8448–4521}}
\author[2]{Suzan Verberne\orcidlink{0000–0002–9609–9505}}

\affil[1]{Centre for Science and Technology Studies, Leiden University, The Netherlands}
\affil[2]{Leiden Institute of Advanced Computer Science, Leiden University, The Netherlands}

\date{December 2024}
\begin{document}

\maketitle
\begin{abstract}
Traditional science maps visualize topics by clustering documents within a network, but they are inherently biased toward clustering certain topics over others. If these topics could be chosen, then the science maps could be tailored for different needs. In this paper, we explore the extent to which the topic bias of a science map can be changed by choosing different data sources to build the document network. We analyze this by evaluating the clustering effectiveness of several topic categories over two sources that are traditionally used for the creation of science maps (citations and text similarity) and six non–traditional data sources, which we found favor different kinds of topics: Health issues for Facebook users, biotechnology topics for patent families, government and social issues for policy documents, food topics for Twitter conversations, nursing topics for Twitter users, and geographical entities for document authors (the favoring in this latter source was particularly strong). Our results show that diverse data sources can be used to control topic bias, which opens up the possibility of creating science maps tailored for different needs.
\end{abstract}

\section{Introduction} \label{Introduction}

Science maps are a form of visualization that provide a content overview of a collection of academic documents. They are typically used for literature analysis \cite{zitt2015meso}, field delimitation, research policy, and enhanced document browsing \cite{bascur2023academic}. A typical practice for creating science maps is first to create a network of academic documents where the links are an aspect of the documents (e.g. bibliographic metadata), then to cluster together the documents that are well connected, and finally to summarize the contents of these clusters. In other words, the map is a set of clusters that emerge from document connections, and what a cluster represents is inferred from its documents. Similar visualization approaches have also been applied to other large–scale knowledge sources, such as Wikipedia \cite{holloway2007analyzing}, while science maps tend to focus on academic production. Science maps originated as a method for visualizing the structure of science production \cite{borner2003visualizing,boyack2005mapping}, and have since also been used for field delimitation and information retrieval. For a comprehensive overview of science mapping and its methodological foundations, we refer to Thijs \cite{thijs2019science}.

The extent to which a clustering solution reflects an underlying ground truth can be assessed using a variety of evaluation metrics, each emphasizing different aspects of agreement between clusters and reference categories. In this work, we focus on clustering effectiveness, a family of measures that captures a trade–off between two objectives: containing the ground–truth documents (or a specified fraction of them) within as few clusters as possible, and minimizing the inclusion of non–ground–truth documents within those clusters \cite{Yuan2022}. This trade–off is directly influenced by the granularity of the clustering: coarser clusterings tend to group ground–truth documents into fewer clusters, while finer clusterings tend to reduce the number of non–ground–truth documents within those clusters. This makes clustering effectiveness particularly relevant for science maps, which are commonly constructed at user–defined levels of granularity.

In our previous work \cite{previouswork}, we used clustering effectiveness to assess the extent to which a science map can place the documents of a given topic into clusters that predominantly contain documents of that same topic; that is, to form clusters about the topic. Our aim was not to evaluate the clustering solution as a whole, but to assess clustering effectiveness at the level of individual topics. This distinction is important because, in science maps, perfect clustering effectiveness cannot be achieved simultaneously for all topics: documents that address multiple topics must still be assigned to a single cluster.

Accordingly, in that earlier study we examined whether there is a systematic bias in which kinds of topics tend to be clustered more effectively. In that study our data sources were citations and text similarity, the latter based on the BM25 score \cite{10.1561/1500000019}. We found that such a bias indeed exists. For example, in maps based on citation links or text similarity, topics related to diseases tend to be clustered much more effectively than topics related to geographical locations. It is worth noting that these comparisons evaluate clustering effectiveness relative to other topics, rather than against an absolute reference that would define what level of quality is sufficient for practical usability. This reflects a broader limitation of evaluation approaches that rely exclusively on automated, ground–truth–based measures. However, alternative evaluation strategies that incorporate human judgment, such as expert–based assessments, have been shown to produce results that are often inconclusive and inconsistent across experts \cite{ahlgren2020enhancing}.

The topic bias can prove inconvenient for science map users if their topics of interest do not align with the topic bias of the map, because then their topics would not be well represented by the map. For example, a science map user that wishes to find research about a given country will find few or no clusters about this country, leading to the wrong conclusion that there is little research about this country. From a user perspective, it is therefore desirable to explore ways to control or adjust topic bias so that science maps can better serve different information needs. In the example above, this would correspond to improving the clustering quality of country–related topics, even if doing so comes at the expense of clustering effectiveness for other kinds of topics. Motivated by this issue, in the current paper we investigate whether the topic bias inherent in commonly used science maps can be altered to better support different kinds of topics. Specifically, we examine whether such changes in topic bias can be achieved by constructing the underlying document networks using alternative data sources.

The traditional data sources to create science maps are citation links networks and text similarity networks, but to achieve our goal we explore other, non–traditional data sources. Most of our non–traditional data sources create networks where two or more academic documents are connected via an element external to the documents themselves (e.g. a patent that cites two documents), and for this reason we will refer to these sources as external sources. Our topics are based on MeSH terms, and we group the topics into topic categories to facilitate our analysis. We measure the topic bias of a network as how well a topic is clustered (i.e. clustering effectiveness) over several clustering solutions, each of them with different cluster sizes. Each of these clustering solutions is analogous to a very simple science map. We use the topic bias of text similarity networks as our reference to compare how the topic bias changes in other networks.

Our research question is: Which topic categories benefit from using each external source? We operationalize this benefit in two ways: First, if the clustering effectiveness of the topic category in the network of the external source is higher than the effectiveness of the same topic category in the text similarity network, and second, if the clustering effectiveness of the topic category in the external source is higher than the other topic categories in the same external source. We will consider both operationalizations to address our research question, but give more importance to the first one because it serves the needs of science map users more directly.

Our contributions are: (1) We present an expanded and improved analysis method for evaluating the clustering effectiveness of a topic; (2) With this method, we provide a large–scale analysis of eight different sources (two traditional and six external), twenty one networks of up to four million documents, nearly three thousand clustering solutions, and seventeen topic categories, each one usually composed of between fifty and three hundred topics (values vary between networks); (3) With this analysis, we show that topic bias can be changed using external sources, and also which topic categories are favored for each of the external source. This knowledge expands the customization options of science maps.

\section{Background}

In this section we explore several topics related to our paper, provide literature examples for each of them, and explore how our paper relates to the most relevant ones.

\subsection{Interaction of academic documents with non–academic elements}

Traditionally, policy makers analyze scientific production to evaluate scientific impact, but they also are interested in evaluating its societal, technological and policy–making impact. For societal impact, the impact of publications on social media has been suggested as a proxy \cite{10.1093/scipol/scac004}, and we highlight the company Altmetric \cite{altmetric,fang2020extensive,dorta2024societal}, which collects mentions to academic documents online, including social media. For technological impact, patents are used \cite{meyer2000does}. Policy–making impact is a more recent field of study, and we highlight the company Overton \cite{szomszor2022overton, fang2024science}, which collects ample datasets of policy documents and their references \cite{dorta2024societal}. We also highlight the company Dimensions \cite{hook2018dimensions}, which collects the connections of academic documents to citations, clinical trials, patents, policy documents, grants and datasets.

\subsection{Science maps based on diverse sources}

Science maps of academic documents typically use networks of citations links or text similarity \cite{waltman2020principled}, but both Janssens, Gl{\"a}nzel, and De Moor \cite{janssens2008hybrid} and Ahlgren et al. \cite{ahlgren2020enhancing} proposed networks that combine both citation links and text similarity. Also, Costas, de Rijcke and Marres \cite{costas2021heterogeneous} proposed a conceptual framework for analyzing the interaction between documents and social media by creating networks of co–occurrence. Their framework is our source of inspiration for using external sources to improve science maps and also for how we build the networks of external sources. The main difference between their networks and our networks is that in their networks co–occurrence is explicitly included in the weight of the edges, while in our networks it is implicit by building the network with both the documents and the elements where the documents co–occur, an approach similar to the work of Yun, Ahn and Lee \cite{yun2020return}. 

An alternative method to create science maps is to create a network where the clusters are not made of academic documents, so to obtain a different perspective on the academic data. Keywords can be used to identify the topics within a collection of documents, connecting the keywords by the documents where they co–occur \cite{lee2021knowledge}. This has a slightly different functionality from identifying topics using document clusters, like to study the evolution of topics over time \cite{wang2024studying}. Authors can be used to identify scientific collaborations, connecting the authors either by their co–authorships \cite{newman2004coauthorship} or their citations \cite{wang2019exploring}. Patents can be used to identify technological developments, connecting the patents by their cited documents \cite{lai2005using}. By their nature, networks of elements that co–occur with academic documents can be turned into networks of documents that co–occur with these elements. For example, Yang and Colavizza \cite{yang2022map} created two networks using the same data, one of documents cited by the same Wikipedia article, and one of Wikipedia articles citing the same document. In this example, the co–occurrences were explicit, but Carusi and Bianchi \cite{carusi2019scientific} created a bipartite network of authors and journals where the co–occurrences were implicit. This allowed them to create clusters for both the authors and the journals using the same network with a method they called co–clustering. In our paper the external source networks are also bipartite, but our methodology will only focus on clustering the academic documents, not the external source elements.

\subsection{Criticisms of maps of science}

There are several criticisms of the capacity of science maps to represent topics. Gläser \cite{glaser2020opening} reported that expert based evaluation of maps is usually inconclusive. Held, Laudel and Gläser \cite{held2021challenges} found that the science maps were unable to have both at the same time one topic per cluster and one cluster per topic. Held and Velden \cite{held2022interpret} found that clusters represent individual species instead of a biological field. Hric, Darst and Fortunato \cite{hric2014community} made a strong criticism of the capacity of any kind of clustering algorithm in any kind of network to create clusters where all the cluster nodes belong to a given category. Because of the failure of science maps to properly cluster all topics, topic wise evaluation of science maps aims to make a more granular evaluation of the clustering and identify which topics get more effectively clustered, instead of making an overall statement about the quality of the map. This area of research has been sparsely explored by the literature. As far as we know, beyond our prior work \cite{previouswork}, the only topical analyses that exist are the expert based evaluations of science maps and, to a lesser extent, the exploration of the epistemic function of intra– and inter–cluster citations performed by Seitz et al. \cite{seitz2021case}.

\subsection{Comparing clustering solutions of different networks}

Different networks generate different science maps, and there have been several attempts to compare the clustering solutions of different networks. Xu et al. \cite{xu2018overlapping} identified overlapping communities between the clusters of two networks with the same nodes. Xie and Waltman \cite{xie2023comparison} did something similar, but using topic modeling instead of text similarity networks. Šubelj, Van Eck and Waltman \cite{vsubelj2016clustering} evaluated the quality of the clusters generated by different clustering algorithms from the same network. Their method evaluated if the topics of the clusters correspond to the topics of the field experts, and also evaluated attributes of the clustering, like clustering stability, computing time, and cluster size. Waltman et al. \cite{waltman2020principled} compared clustering solutions from different networks with the same nodes using an additional network as reference to calculate the accuracy of the clusters. For an example that does not use clustering, Ba and Liang \cite{ba2021novel} identified overlapping edges between two networks with the same nodes. In our prior work \cite{previouswork}, we compared the clustering effectiveness per topic by evaluating the extent to which topic documents are in few clusters and the extent to which these same clusters only contain topic documents. In the current paper we refine this method so its results are easier to interpret.

\section{Methods}

In this section we describe how we obtained and cleaned the data, created the networks and clusters, evaluated the clustering effectiveness, and compared the topic categories. The data and code used are available in the Supplementary material Section.

\subsection{Core academic documents}

This is the set of documents that we used in the evaluation of clustering effectiveness, and each network has a different subset of these documents depending on the data available for each external source. We selected all Web of Science documents from the CWTS local database published between the years 2016 and 2019 that have a PubMed id (which is necessary to have MeSH terms) and that have a noun phrase in the title or abstract sections. The latter condition was added to have high quality text similarity networks, and the noun phrases were identified using the method developed by Waltman and van Eck \cite{waltman2012new}. We chose this range of years so as to have enough connections between the documents and the external source elements, especially with patents because they take multiple years to accumulate, and also because in these years Twitter became popular for sharing academic documents while not being the years of the Coronavirus pandemic. The external source elements are not limited by this time period. In total, our core set contains 4,142,511 documents.

\subsection{External sources networks}

The external source networks are built the following way: For each external source, we first define what the nodes of this source mean (e.g. academic document authors, facebook users, etc…), which we will refer to as the external source “elements”. Then we select core academic documents and external source elements that we will use in the network, such that all the documents are connected to at least one element and all the elements are connected to at least two documents. We use the “at least two documents” threshold so that we do not have documents without any indirect connections with other documents (there are no direct connections between documents). Then we create a network with these documents and elements where the edges that connect them are undirected and have weight value 1, the document nodes have weight value 1 and the element nodes have weight value 0. We give this weight value to the element nodes so that the clustering algorithm does not take these nodes into account when calculating the quality of a cluster. We will refer to these networks as the “Pure” networks of an external source, to distinguish them from the mixed and the text similarity networks of an external source (see below). It is worth mentioning that this network creation design creates a bipartite network (only document to non–document edges), while in science mapping literature it is more common to represent these relations as a co–occurrence network (only document to document edges with no non–document nodes, and the weight value of the edge is the number of non–document elements in common between the documents). We use bipartite networks because they represent these relations with more computational efficiency than co–occurrence networks.

We used the following external sources. All databases are CWTS 2023 local snapshots, and the external source elements we selected were published between 2016 and 2023.

\textbf{Documents authors (AUTHOR):} The external source elements are the authors of academic documents, and the connections are to these documents. The data comes from the disambiguated authors database of CWTS \cite{d2020collecting}. This network has 3,977,303 core academic documents, 2,710,012 external source elements and 19,820,564 edges.

\textbf{Facebook users (FACEBOOK):} The external source elements are the Facebook users (i.e. accounts), and the connections are to the documents they have posted web links to. The data comes from the Altmetric \cite{altmetric} Facebook database. This network has 596,783 core academic documents, 44,811 external source elements and 1,231,887 edges.

\textbf{Twitter users (TWUSER):} The external source elements are the Twitter users (i.e. accounts), and the connections are to the documents that their tweets have web links to. The data comes from the Altmetric \cite{altmetric} Twitter database. This network has 2,364,304 core academic documents, 1,495,275 external source elements and 27,981,494 edges.

\textbf{Twitter conversations (TWCONV):} The external source elements are the Twitter conversations, and the connections are to the documents that its tweets have web links to. A Twitter conversation is an original (non–reply) tweet plus all the tweets that directly or indirectly reply to it. The data comes from the Altmetric \cite{altmetric} Twitter database. This network has 227,212 core academic documents, 493,049 external source elements and 1,175,624 edges.

\textbf{Patent families (PATENT):} The external source elements are patent families, and the connections are to the documents cited by the patents of the patent family. A patent family is made up of an initially submitted patent, plus derivative patents (like updates or new application) and versions of the patent submitted in different countries. The data comes from the PATSTAT database \cite{kang2016patstat} and we only use invention patents. This network has 98,278 core academic documents, 41,714 external source elements and 175,693 edges.

\textbf{Policy documents (POLICY):} The external source elements are policy documents, and the connections are to the documents cited by the policy documents. A policy document is a document written primarily for policy makers, and includes documents such as memos and guidelines from governments and think tanks. The data comes from the Overton database \cite{szomszor2022overton}. This network has 311,867 core academic documents, 64,951 external source elements and 651,099 edges.

\subsection{Text similarity networks}

We use the topic bias of text similarity networks in our experiments as a reference to compare how the topic bias changes in other networks. We chose this source because it is traditionally used for the creation of science maps and also because it is less computationally demanding to create and cluster than the citation network, which is relevant because we created a reference network for each external source.

The method to measure text similarity is the cosine similarity between the BERT–based embeddings of the text of two documents. Unlike other text similarity metrics, this measurement does not rely on explicit word overlap. The text of a document is its concatenated title and abstract, and the embedding of the text is extracted using the Python implementation of Sentence–BERT \cite{reimers2019sentence} with the model allenai–specter. Sentence–BERT is a software that runs BERT–based models, but unlike standard BERT, its embeddings are designed to reflect semantic similarity through cosine similarity. The model allenai–specter provides access to the BERT–based model ``SPECTER'' \cite{cohan2020specter}, which is a model optimized to create embeddings of the title and abstract of academic documents.

For each external source, we create a text similarity network that contains the same academic core documents as the Pure network, which we will refer to as the “BERT” network, and we also create a network that combines both networks, which we will refer to as “Mixed” network. To create the BERT network of a source we first make the academic documents into nodes with weight value 1. Then, we calculate the text similarity between all pairs of documents and only keep the 20 highest pairs per document. These values become the weights of the undirected edges between the nodes, and if there are two edges between two nodes then we merge them and sum their weights. Finally, we multiply all the edge weight values by a factor such that the sum of all edge weight values in a network is the same for the BERT and the Pure networks. To create the Mixed network of a source we use the Pure network and add to it the edges from the BERT network. The purpose of the step where we multiply the edge weight values by a factor is to bring this network to the same magnitude as the Pure network, which has two goals: To make the edges that came from the BERT and Pure network have the same magnitude of influence in the clustering of the Mixed network, and to use the same clustering Resolution values for the BERT and Pure networks, which is just convenient.

\subsection{Citation network}

There are not many science maps studies published using Sentence–BERT for text similarity because it is a recently developed method, making our results difficult to compare to the literature. To solve this, we also evaluated the topic bias of a network that is built based on a method well researched in the literature and presented it next to the other external source networks. This well published method is the extended direct citation \cite{waltman2020principled}, which is a citation network that includes connections to academic documents that are not part of the core academic documents. The Pure citation network includes all the core academic documents as nodes with weight value 1 and the citations between each other as undirected edges with weight value 1. It also includes the non–core documents from Web of Science that have citation links to at least two core academic documents as nodes with weight value 0, and these links as undirected edges with weight value 1. These non–core documents are usually documents from outside the time period or that do not have a PubMed id. This network has 4,142,511 core academic documents, 18,960,516 non–core academic documents and 217,907,980 edges. The Mixed and BERT citation networks are created the same way as for the Mixed and BERT of the external sources.

\subsection{Clustering}

To cluster we used the Leiden algorithm \cite{traag2019louvain}, which is typically used in science maps. This algorithm requires the user to set a parameter, the “Resolution”, which has an effect on the size of the clusters (higher Resolution, smaller clusters). We clustered each network several times using a wide range of Resolution values, using a different value each time. We decided on the Resolution values range on a network wise basis, and our criteria for this range was for the highest value to create a clustering solution where most clusters have only one node, and for the lowest value to create a clustering solution where most of the nodes belong to a single cluster. We clustered a number of Resolution values that allowed us to keep the running time manageable (between 70 and 140 Resolution values per network), using the Python implementation of the library Igraph \cite{igraph} and the Leiden algorithm. All the clustering solutions are used during the evaluations and comparisons.

\subsection{Topics and topic categories} \label{Topics and topic categories}

Our topics are the tree nodes in the MeSH hierarchical tree of MeSH terms, and the topic documents of a given topic are the documents labeled with the tree node of a topic. MeSH terms are a controlled vocabulary thesaurus from the National Library of Medicine (NLM) used for indexing PubMed, and are semi–automatically annotated to documents by the NLM \cite{mesh}. We use MeSH terms instead of other alternatives because of their extensive system of hierarchical topics, high number of annotated documents, and high quality of annotations. The MeSH terms are organized in a hierarchical tree where almost each MeSH term maps to one or more nodes in the tree, but each tree node maps to a single MeSH term. The tree is composed of 16 branches, and the tree nodes in the lower levels are subtopics of the tree nodes in the higher levels. We refer to a tree node using their MeSH term name followed by their tree node identity (e.g \textit{Head [A01.456]}). The reason why we base our topics on the tree nodes of the MeSH terms instead of just using the MeSH terms themselves is to facilitate the expansion and filtering of topics in the next steps of the methodology (see below). We obtained the MeSH terms annotated for each document, plus the metadata of the MeSH terms themselves, including their tree nodes, from the in–house CWTS database of PubMed and MeSH (version from 2024).

Our topic categories are the MeSH tree branches, and all the tree nodes in the branch are topics that belong to the topic category. We use branches as topic categories because they are epistemic categories (e.g., organisms), which are the kind categories commonly used for topical analysis of clusters \cite{seitz2021case,previouswork}.  There are 3 branches that we decided to, instead of using them as topic categories, use their highest level tree nodes as topic categories, because we think these tree nodes work better than their branches as topic categories. The branches that we replaced with their higher level tree nodes are \textit{Disciplines and Occupations [H]}, \textit{Anthropology, Education, Sociology, and Social Phenomena [I]} and \textit{Technology, Industry, and Agriculture [J]}. We also removed the following topic categories due to having too few topics: \textit{Humanities [K]}, \textit{Publication Characteristics [V]}, \textit{Human Activities [I03]}, and \textit{Non–Medical Public and Private Facilities [J03]}. In the end, we used the 17 topic categories in Table \ref{table:topic_categories}.

\begin{table}[htbp]
\centering
\caption{List of topic categories used in the current paper.}
\begin{tabular}{c}
\hline
\textbf{Topic Categories} \\
\hline
Anatomy [A] \\
Organisms [B] \\
Diseases [C] \\
Chemicals and Drugs [D] \\
Analytical, Diagnostic and Therapeutic Techniques, and Equipment [E] \\
Psychiatry and Psychology [F] \\
Phenomena and Processes [G] \\
Natural Science Disciplines [H01] \\
Health Occupations [H02] \\
Social Sciences [I01] \\
Education [I02] \\
Technology, Industry, and Agriculture [J01] \\
Food and Beverages [J02] \\
Information Science [L] \\
Named Groups [M] \\
Health Care [N] \\
Geographicals [Z] \\
\hline
\end{tabular}
\label{table:topic_categories}
\end{table}

To have good topics, we would like each topic to be annotated on all the documents related to it, but the NLM typically only annotates up to fifteen MeSH terms per document, which means that the more generic MeSH terms are not annotated. To fix this, we expanded the topics annotated on a document using the already annotated MeSH terms and the MeSH tree. We transformed each of the MeSH terms into all of their corresponding MeSH tree nodes, and then we added all the MeSH tree nodes upstream in the MeSH tree from the current MeSH tree nodes. For example, if a document had the MeSH term \textit{Scalp}, we transformed this MeSH term into its tree node version (\textit{Scalp [A01.456.810]}), and added the upstream tree nodes (\textit{Head [A01.456]}, \textit{Body Regions [A01]}) to the document.

To improve the reliability of our evaluation we filter our topics. We do this filtering process for each external source because they use different sets of core academic documents. Our first filter criterion is by topic size (i.e. number of documents with the topic) because the size of a topic can affect its clustering effectiveness. We group the topics by size into Size bins, which go from a value (excluding it) to double that value (including it), starting at 40 (e.g. 41–80, 81–160, 161–320, … $[X+1]$–$[2X]$). We use 40 for reasons explained in Section \ref{Clustering effectiveness}. We filter out the Size bins that have less than half the number of topics than the Size bin with most topics, and also filter out the topics that belonged to these filtered out Size bins. The Size bins that we keep per source are shown in Table \ref{table:size_bins}.

\begin{table}[htbp]
\centering
\caption{Size bins per source after filtering.}
\begin{tabular}{ll}
\hline
\textbf{Source} & \textbf{Size Bins} \\
\hline
Patent families & 41–80; 81–160; 161–320 \\
Policy documents & 41–80; 81–160; 161–320 \\
Facebook users & 41–80; 81–160; 161–320; 321–640 \\
Twitter conversations & 41–80; 81–160; 161–320; 321–640 \\
Twitter users & 81–160; 161–320; 321–640; 641–1,280 \\
Documents authors & 161–320; 321–640; 641–1,280; 1,281–2,560 \\
Citations& 161–320; 321–640; 641–1,280; 1,281–2,560 \\
\hline
\end{tabular}
\label{table:size_bins}
\end{table}

Our second filter criterion is redundancy (i.e. two topics share a substantial number of documents) because it can distort our results. To filter by redundancy, we first identify the topics within the same topic category that are redundant with each other. We define two topics as being redundant if they have a Jaccard similarity of 0.5 or higher (calculated from their number of shared documents). We group the redundant topics using the agglomerative hierarchical clustering algorithm with the Complete Linkage method \citation{aggclus} and Jaccard distance, with 0.5 as threshold. Then, we filter out each but the smallest topic from each group, which in our experience tends to also be the topic that best represents the other topics in the group. For example, if there is a group of redundant topics made up of \textit{Canidae [B01.050.150.900.649.313.750.250.216]} and \textit{Dogs [B01.050.150.900.649.313.750.250.216.200]}, we believe that these topics are better represented by the latter than the former. In cases where a group had more than one smallest topic, we selected the one with the tree node at the lowest level in the tree. After filtering topics, we also filter the topic categories that contain too few topics in a Size bin. We chose this threshold manually per external source, but it is always at least between 5 and 10 topics. It is worth mentioning that in our prior work \cite{previouswork} we defined two topics as being redundant if they had Jaccard similarity 0.9 or higher, so in the current paper we are being substantially stricter at ensuring the quality of the data.

\subsection{Evaluation}

\subsubsection{Clustering effectiveness} \label{Clustering effectiveness}

To find out which topics are better represented by the clustering of the networks, we use the concept of clustering effectiveness that we introduced in our prior work \cite{previouswork}. The unit to measure the clustering effectiveness is “Purity”, which is, for a set of selected clusters, the fraction of their documents that belong to a given topic. In mathematical terms, Purity is defined as:
 \begin{equation}
 Purity = \frac{\sum_{i=1}^{N}{|D_i \cap D_{M}|}}{\sum_{i=1}^{N}|D_i|}
 \label{equation:purity}
 \end{equation}
Here, $N$ denotes the number of selected clusters, $D_i$ denotes the documents in selected cluster $i$ and $D_M$ denotes the topic documents of the topic (these concepts are clarified below). The higher Purity, the more effective the clustering. Purity is bounded between values zero and one, with Purity value one meaning that the selected clusters only contain topic documents. 

We calculate Purity for each clustering solution and topic, but instead of selecting all the clusters that contain topic documents to calculate Purity, we only select a subset of these clusters. To do this, we sort all the clusters that contain topic documents from the highest to the lowest number of topic documents, with ties won by the smallest cluster. Then, we choose the threshold of the minimum number of topic documents that we want the set of selected clusters to contain, and then select clusters in the sorted order until we reach this threshold. We call this value Coverage, and it is a fraction of the total number of topic documents. In our paper we calculate Purity for three Coverage values: 0.25, 0.50 and 0.75. We only compare Purity values calculated using the same Coverage value. In reference to Section \ref{Topics and topic categories}, the reason why Size bins start at 40 is because at Coverage 0.25 the value of the threshold is only 10 documents, which we set as the minimum to have a meaningful academic topic.

To illustrate this procedure, consider the computation of Purity for topic $M$ in a clustering solution $Sol$ at Coverage $Cov$. Let $|D_M|$ denote the total number of documents associated with topic $M$. The threshold $t$ for the minimum number of topic documents is defined as $t = |D_M| \cdot Cov$. All clusters in $Sol$ that contain at least one topic document are sorted in descending order by $|D_i \cap D_M|$, with ties broken by ascending cluster size $|D_i|$. Clusters are selected sequentially until the cumulative number of topic documents in the selected clusters is greater than or equal to $t$. This set of clusters is then used to compute Purity. As a concrete example, suppose that three clusters are selected with sizes $|D_i| = 100$, $200$, and $300$, and that each contains $|D_i \cap D_M| = 10$ topic documents. In this case, Purity is given by
\begin{equation}
\mathrm{Purity} = \frac{10 + 10 + 10}{100 + 200 + 300} = 0.05
\end{equation}

In our concept of clustering effectiveness, the number of selected clusters (NSC) also plays a role. In a science map, finding clusters related to a topic requires effort, so the smaller the NSC, the higher the cluster effectiveness. Also, a high NSC is correlated with smaller clusters, which itself is correlated with higher Purity because smaller clusters allow a more fine selection of the clusters. For example, if all clusters in a clustering solution are size one, then the value of Purity is also one because all the selected clusters contain only topic documents. To control for the effect of NSC over Purity, we only compare Purity values when they have the same NSC. 

\subsubsection{Topic Purity profiles}

\begin{figure}[t]
 \centering
 \includegraphics[width=0.50\columnwidth]{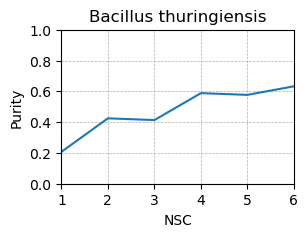}
 \caption{Example of a Purity profile. This is the Purity profile of the topic \textit{Bacillus thuringiensis [B03.510.460.410.158.218.800]} for the Policy documents BERT network calculated using Coverage 0.50. This topic has 60 topic documents among the core documents used by the Policy networks, which for this Coverage value means that the Purity is calculated after selecting clusters that contain at least 30 topic documents. So for example, if we assume that the selected clusters contain exactly 30 topic documents, from the figure we can say that at different Resolution values the network can place 30 out the 60 topic documents in one cluster containing 150 documents ($30/0.2$), two clusters containing 75 documents ($30/0.4$), and four clusters containing 50 documents ($30/0.6$). Using lower Coverage values or topics with more topic documents tends to achieve higher Purity by the last NSC value.}
 \label{figure:Example of a Purity profile.}
\end{figure}

In our research question, we operationalized the concept of "benefit" in two ways, and in the current Section we will present our results such to address both ways. The first operationalization was if the clustering effectiveness of the topic category in the external source (either the Pure or Mixed network) is higher than the same topic category in text similarity (the BERT network). We answer this question by comparing the clustering effectiveness of each topic between these networks, and we represent the clustering effectiveness of a topic for a given network as a series of Purity and NSC values that we will refer to as the topic “Purity profile”. The different Purity and NSC values come from each of the clustering solutions generated for the network. The Purity profile of a topic is the Purity for each NSC value, the NSC values are a consecutive sequence of integers that go from $1$ to $N$ , and $N$ is:
 \begin{equation}
 N = \lfloor\frac{|D_M|*Cov}{5}\rfloor
 \label{equation:profile}
 \end{equation}
Here, $|D_M|$ is the number of topic documents (i.e. topic size), $Cov$ is the coverage value, and function $\lfloor x \rfloor$ means rounded down to the nearest integer. Therefore, the number of NSC values in a Purity profile depends on the size of the topic. The denominator value five ensures that the average number of topic documents per selected cluster is five or more. This average number of documents was chosen based on practical judgment with the purpose of suppressing evaluations that are not meaningful for science maps, such as selecting clusters with a single topic document. As for the Purity values, if there is more than one Purity value for a given NSC, we only use the highest one. If there is no Purity value for the NSC value one, we use Purity value zero. If there is any other Purity value missing for a given NSC value, we estimate it by linear interpolation between the Purity values of the two nearest NSC values with known Purity. Figure \ref{figure:Example of a Purity profile.} is an example of how the Purity profile of a topic looks like.

To compare two topic Purity profiles, we compare their Purity at each NSC value (they must have the same NSC values), and say that one Purity profile is higher than another if they have a higher Purity in at least half of their NSC values. Figure \ref{figure:Diagram on the representation of results.}A shows an example diagram of how we calculate these results. We refer to the fraction of topics in a topic category that are higher than in BERT as the “absolute Purity difference” of this topic category. This value answers the first operationalization of our research question, and it indicates the extent to which the topics of a topic category achieve higher Purity in the Pure or Mixed network than in the BERT network. For example, if the absolute Purity difference of a topic category in the Pure network of an external source is 0.25, it means that a quarter of its topics have higher Purity in the Pure network than the BERT network.

\subsubsection{Topic category Purity profiles}

The second operationalization was if the clustering effectiveness of the topic category in the external source (either the Pure, Mixed or BERT network) is higher than the other topic categories in the same external source. To compare topic categories within a network we create a topic category Purity profile for each of their Size bins, and only compare profiles with the same Size bin. To create the Purity profile of a topic category at a given Size bin, we first obtain its Purity and NSC values for each Resolution. However, Purity and NSC values are only defined for topics, not for topic categories, so we define the Purity and NSC values of a topic category at a given Size bin and Resolution values as the median Purity and NSC of the topics of that topic category in that Size bin and at that Resolution. 

Then, to build the topic category Purity profile, we use the Purity and NSC values following the same protocol that we did for building the Purity profiles of topics, with the difference that to calculate the $N$ of the Purity profile (i.e. the highest NSC value in the profile) in Equation \ref{equation:profile} we replace $|D_M|$ with $S$, which is the average between the lower and upper bound of a Size bin (e.g. for Size bin 41–80, $S=60$, and if $Cov=0.25$, then $N=3$). 

As a note, we want to mention that we considered using topic category Purity profiles instead of topic Purity profiles for the first operationalization, but we found that the results from this approach provided us with less nuanced information than the one we ultimately used. However, as a general rule, if half or more of the topic Purity profiles in a topic category were higher in the external source than in BERT (absolute Purity difference $> 0.5$), then their different Size bin topic category Purity profiles were also higher than in BERT.

\begin{figure}[t]
 \centering
 \includegraphics[width=0.90\columnwidth]{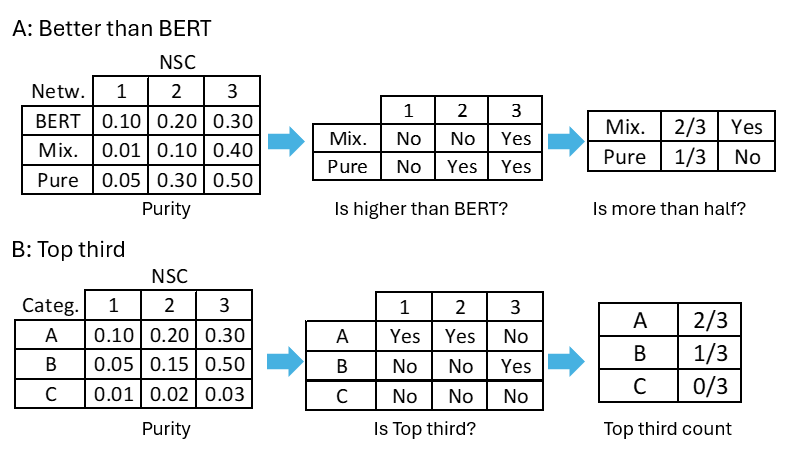}
 \caption{Diagram on the representation of results. A: How to calculate from topic Purity profiles if a topic has higher clustering effectiveness than BERT in the Pure or the Mixed network. In this example, a topic has higher Purity than BERT for the Mixed network, but not so for the Pure network. B: How to calculate from topic category Purity profiles the number of NSC that a topic category is in the top third Purity of a network. In this example, the topic categories A, B and C achieve a top third count of 0.7, 0.3 and 0, respectively}
 \label{figure:Diagram on the representation of results.}
\end{figure}

To compare the topic categories Purity profiles of a given network with each other, we do not calculate which one is higher as we did for the topic Purity profiles, but instead we calculate how often their Purity is higher than each other. We do this because we want to consider all the topic categories at the same time, and also because which one is better tends to change frequently across the NSC, probably because some of them have very similar clustering effectiveness. Therefore, for each Size bin in a network, we consider all the topic categories Purity profiles at the same time, and for each NSC value (they all have the same NSC values because they have the same Size bin and Coverage) we identify which topic categories are among the top third highest Purity value at this NSC value. Then, for each topic category we report for how many (as a fraction) NSC values it is among the top third, averaged over all the Size bins. Figure \ref{figure:Diagram on the representation of results.}B shows a diagram of how we calculate these results. For example, if the top third count of a topic category in the Pure network of an external source is 0.25, it means that, on average across the Size bins, it is among the top third highest Purity topic categories of the Pure network for a quarter of the NSC. We defined the top value in relative terms (as a third) because different external sources have a different number of topic categories. 

The top third count already answers the second operationalization, but we would like to go one step further to know how the external source is different from text similarity. To do this, we subtract the Pure or Mixed network top third count from the BERT top third count to obtain a value that we refer to as the “relative Purity difference” of that topic category. This value indicates the extent to which a topic category achieves higher Purity than the other topic categories in the Pure or Mixed networks, but not so in the BERT network. For example, if the relative Purity difference of a topic category in the Pure network of an external source is 0.25, it means that in the Pure network the top third count of that topic category is 0.25 higher than in the BERT network.

\subsubsection{Summary of comparisons methods}

We compare the clustering effectiveness of topic categories for an external source using two metrics: The absolute Purity difference, which indicates the difference between the Pure or Mixed networks against the BERT network for this topic category, and relative Purity difference, which indicates the difference of how this topic category relates to the other topic categories within the Pure or Mixed networks against how it relates in the BERT network. Both the absolute and relative Purity differences are important to understand which topic categories benefit from using each external source. The absolute difference is the most important because it indicates if a topic does better or worse than in BERT, while the relative difference only ranks topic categories within the network. However, the Purity could increase with better methods for creating the clusters and networks, especially considering that we did not focus on achieving high Purity. In such cases, the relative difference can suggest which topic categories can achieve high Purity after refining the methods, even if we achieved a low absolute difference.

\section{Results} \label{Results}

\begin{figure}[htbp]
 \centering
 \includegraphics[width=0.30\columnwidth]{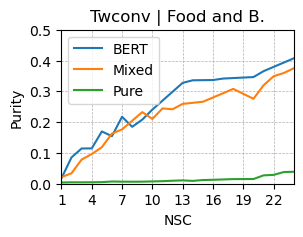}
 \includegraphics[width=0.30\columnwidth]{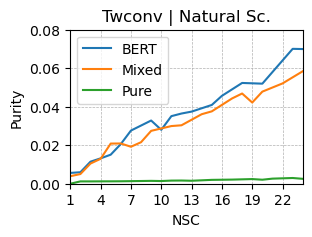}
 \includegraphics[width=0.30\columnwidth]{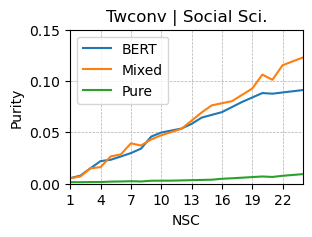}
 \includegraphics[width=0.30\columnwidth]{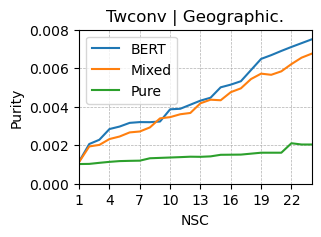}
 \includegraphics[width=0.30\columnwidth]{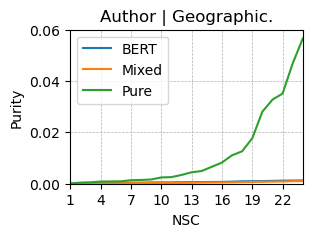}
 \includegraphics[width=0.30\columnwidth]{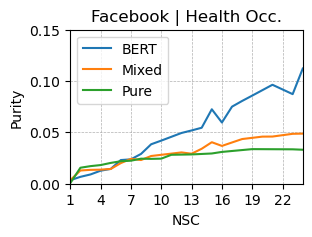}
 \includegraphics[width=0.30\columnwidth]{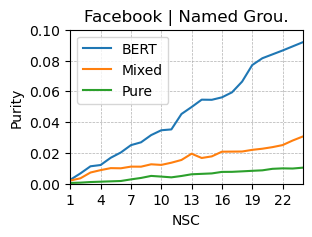}
 \includegraphics[width=0.30\columnwidth]{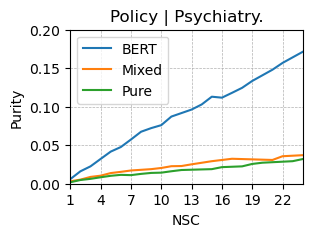}
 \includegraphics[width=0.30\columnwidth]{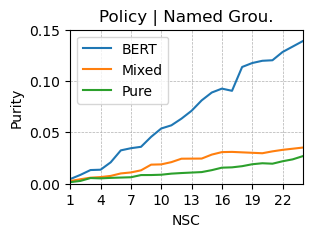}
 \includegraphics[width=0.30\columnwidth]{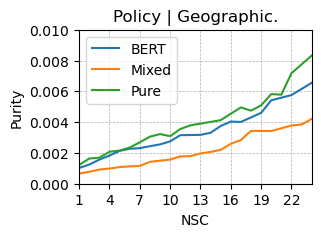}
 \includegraphics[width=0.30\columnwidth]{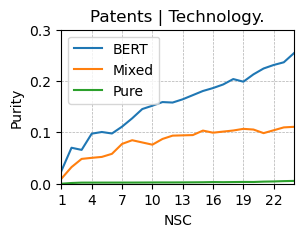}
 \includegraphics[width=0.30\columnwidth]{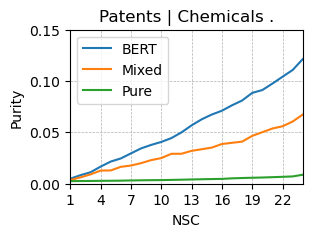}
 \includegraphics[width=0.30\columnwidth]{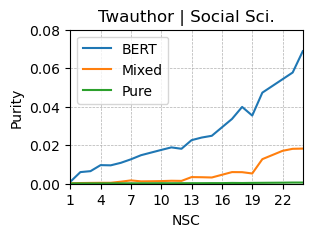}
 \includegraphics[width=0.30\columnwidth]{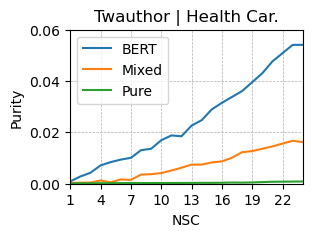}
 \caption{Examples of Purity of several topic categories for different networks. These examples are referenced in Section \ref{Results}. All Purity profiles are calculated for Size bin 161–320 and Coverage 0.50. It is worth reminding that the Purity profiles of topic categories are made of the median Purity and NSC of the topics of that Size bin and topic category at each of the clustering solutions. However, a way to interpret them is to imagine they are the Purity profile of a topic that represents all topics of the topic category. This imaginary topic would have the average size of the Size bin (240 for Size bin 161–320), and therefore each NSC would contain at least 120 topic documents. It is important to mention that, even as they all have the same Size bin, the Purity values should not be compared between different external sources because each source has a different set of core academic documents. This is why we only compare networks that belong to the same external source, as they have the same set of core documents.}
 \label{figure:Examples of Purity profiles of several topic categories.}
\end{figure}

From now on, we will refer to specific networks of an external source using the following prefixes: “b” for the BERT network, “m” for the Mixed network, and “p” for the Pure network, so for example “mTwconv” is the Mixed network of the Twitter conversations. In the current Section we will present the results of our experiments (which are reported in Table \ref{figure:Detail of the results of each network.} and summarized in Table \ref{figure:Summary of the results for each network.}). We will review these results going over each external source, focused on which networks did the best per topic category and the magnitude of this performance, which is also summarized in Table \ref{table:best_networks}. We will limit our discussion of the topic category  \textit{Organisms [B]} because most external sources showed an improvement on it, which suggests that BERT is particularly bad at it. We will also limit our discussion of the Coverage because the three Coverage values produced roughly the same result, with very few exceptions. For the topic categories that look interesting, we explore if there is a common theme among their high Purity topics. We also explore their topic category Purity profiles (Figure \ref{figure:Examples of Purity profiles of several topic categories.}) to see if they are “competitive”, which means that its Purity profile is close or higher to BERT, and therefore, a science map created using this network might achieve a clustering effectiveness similar or higher than a BERT network for this topic category.

\begin{table}[htbp]
 \centering
 \includegraphics[width=0.45\columnwidth]{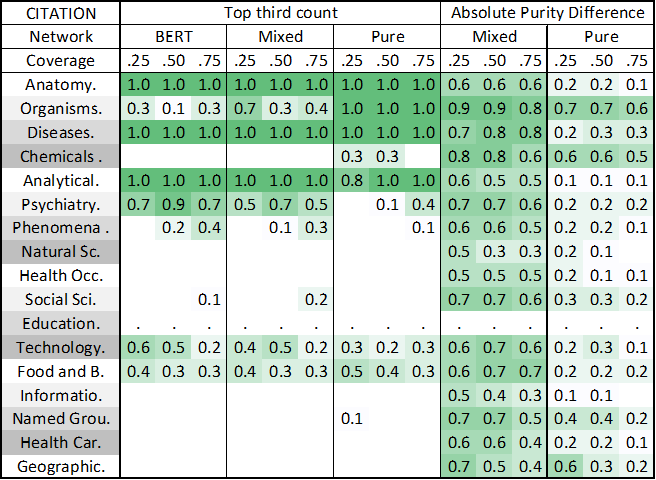}
  \includegraphics[width=0.45\columnwidth]{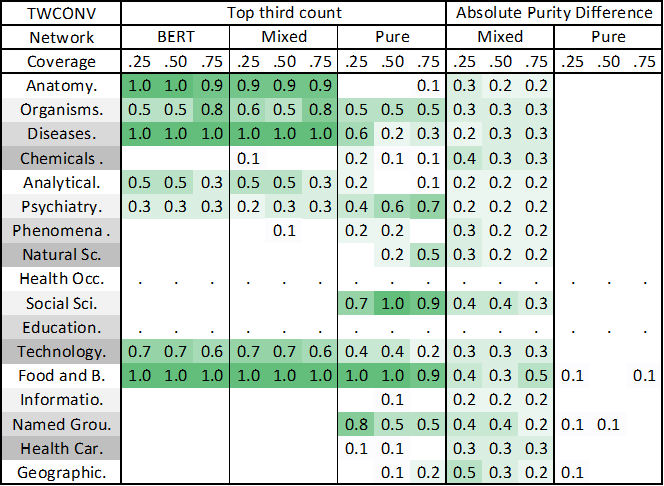}
  \includegraphics[width=0.45\columnwidth]{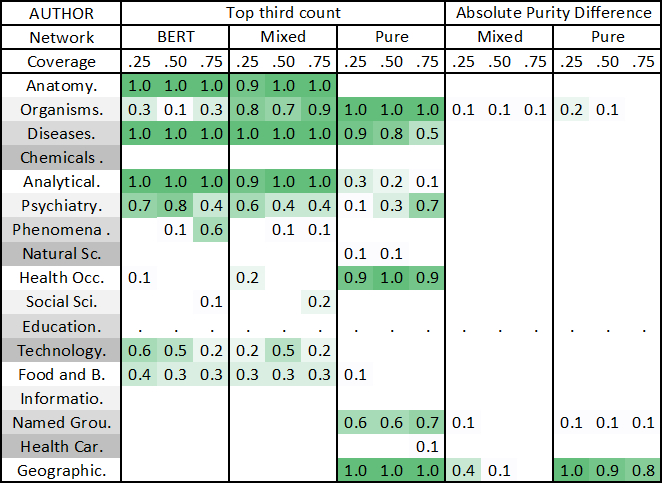}
  \includegraphics[width=0.45\columnwidth]{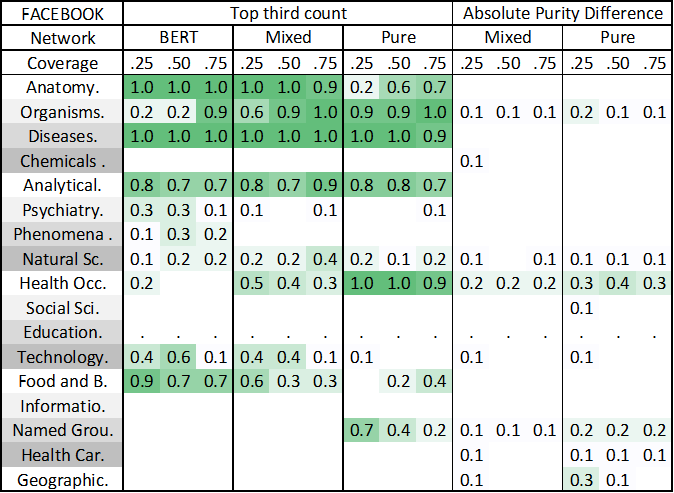}
  \includegraphics[width=0.45\columnwidth]{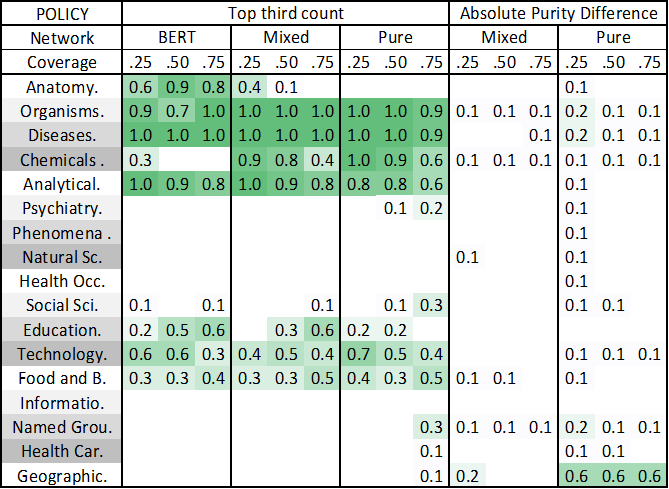}
  \includegraphics[width=0.45\columnwidth]{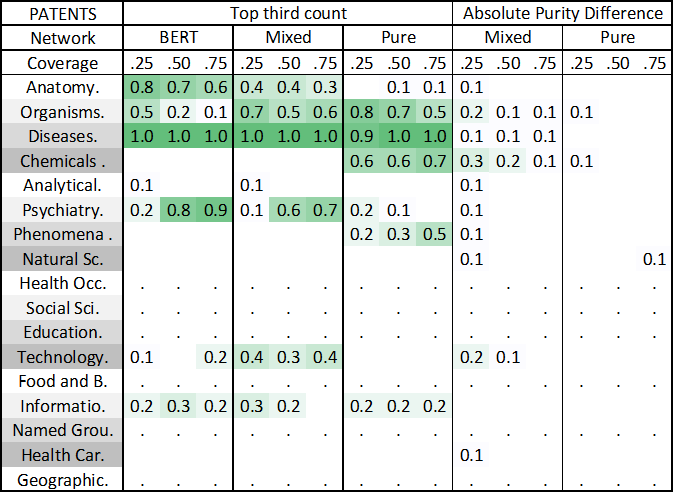}
  \includegraphics[width=0.45\columnwidth]{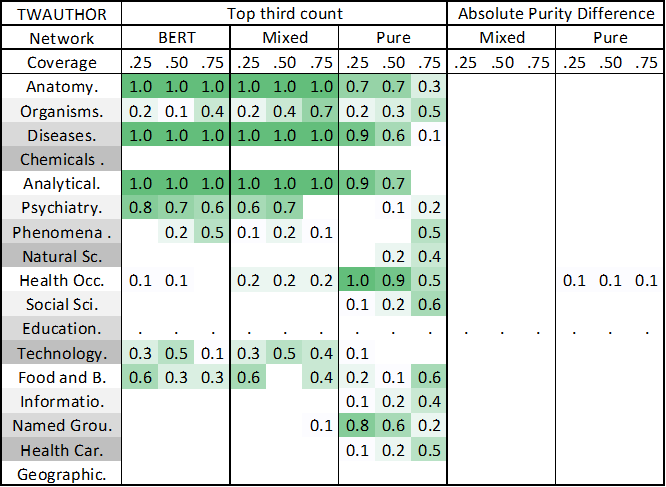}
 \caption{Detail of the results of each network. For each topic category, we show the top third count and the absolute Purity difference at each Coverage value. Values zero are not shown. Dots mean that the topic category was not included in the experiment due to having too few topics per Size bin, as explained in the filtering process.}
 \label{figure:Detail of the results of each network.}
\end{table}
\begin{table}[htbp]
 \centering
 \includegraphics[width=0.90\columnwidth]{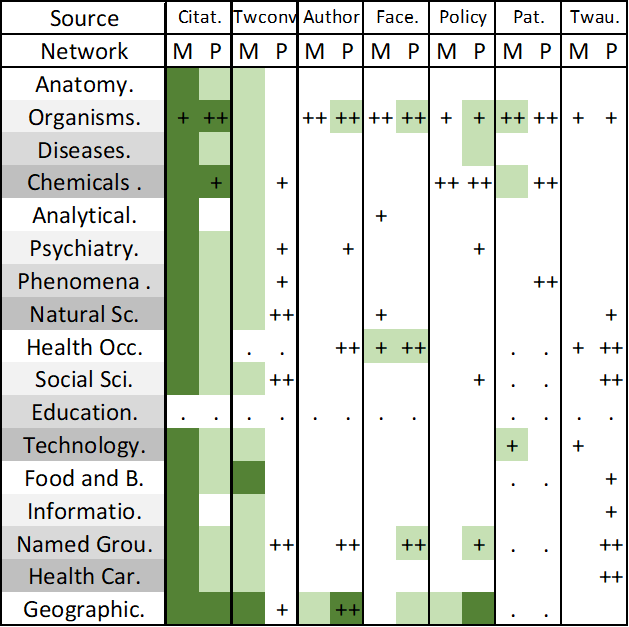}
 \caption{Summary of the results for each network. This table shows the absolute and relative Purity difference, but only the highest of the three Coverage values. All values are derived from Table \ref{figure:Detail of the results of each network.}. “M” means the Mixed network and “P” means the Pure network. Light green and dark green means that the topic category has at least 0.2 and 0.5 absolute Purity difference, respectively. One and two plus signs mean that the topic category has at least 0.2 and 0.5 relative Purity difference, respectively. The relative Purity difference is calculated from the top third count in Table \ref{figure:Detail of the results of each network.}. Dots mean that the topic category was not included in the experiment due to having too few topics per Size bin, as explained in the filtering process.}
 \label{figure:Summary of the results for each network.}
\end{table}
\begin{table}[htbp]
\centering
\caption{Best networks per topic category from Table \ref{figure:Summary of the results for each network.}. We selected the network where the topic category had the highest absolute difference, relative difference, or a combination of both if possible. We gave more importance to absolute difference than to relative difference (i.e., in Table \ref{figure:Summary of the results for each network.}, dark green is better than two plus symbols) because it is more important for our research question. The magnitude, shown by a number of stars, indicates the value of the best networks differences: Zero stars means either one plus symbol, two plus symbols, or light green. One star means both one plus symbol and light green. Two stars mean either dark green or light green with two plus symbols. Three stars mean dark green with two plus symbols. In the abbreviations, “b” stands for the BERT network, “m” for the Mixed network, and “p” for the Pure network. ‘Tw’ stands for Twitter, and ‘conv’ stands for conversations.}
\begin{tabular}{lll}
\hline
\textbf{Category} & \textbf{Best Networks} & \textbf{Magnitude} \\
\hline
Anatomy & mTwconv & \\
Organisms & mPatents, pFacebook, pAuthor & ** \\
Diseases & pPolicy, mTwconv & \\
Chemicals & mPatents, pPatents, mPolicy, pPolicy, mTwconv & \\
Analytical & mFacebook, mTwconv & \\
Psychiatry & pPolicy, mTwconv, pTwconv, pAuthor & \\
Phenomena & pPatents, mTwconv & \\
Natural Sc. & mTwconv, pTwconv & \\
Health Occ. & pFacebook & ** \\
Social Sci. & mTwconv, pTwconv, pTwauthor & \\
Education & – & \\
Technology & mPatents & * \\
Food and B. & mTwconv & ** \\
Informatio. & mTwconv, pTwauthor & \\
Named Grou. & pFacebook & ** \\
Health Car. & mTwconv, pTwauthor & \\
Geographic & pAuthor & *** \\
\hline
\end{tabular}
\label{table:best_networks}
\end{table}

\subsection{Citations}

As Table \ref{figure:Summary of the results for each network.} shows, mCitation did better than BERT, and it was the best network overall, which aligned with prior results in the literature where networks that combine citations and text outperform both \cite{boyack2020comparison,waltman2020principled}. pCitation also did better than the other external sources, especially with \textit{Chemicals and Drugs [D]},  but for most topic categories it did not match BERT performance (i.e. absolute Purity difference $< 0.5$), which shows that BERT is a good reference source for our analysis (with the exception of the topic category \textit{Organisms [B]}). The performance of BERT over pCitation is also interesting because in our prior work \cite{previouswork} we compared citation networks (using the same network creating method) with text similarity networks based on the BM25 text similarity metric (a metric that matches and weights the words in common between documents), and found that they had similar clustering effectiveness. This suggests that BERT does better than BM25, which makes sense because BERT is a much more refined method, but we did not test this comparison directly. The fact that most networks outperform BERT at \textit{Organisms [B]} could be because BERT is an embedding representation of the text, which means that it learns the context of words. Given that the words that surround the name of organisms can be very similar, BERT might struggle distinguishing them, so for this topic category it might be better to do text vectorization by term frequency (like BM25) instead of embeddings.

\subsection{Twitter conversations}

The network mTwconv had the best overall performance (after citation networks) because it has at least 0.2 absolute Purity difference in every category. We believe this happens because conversations are more topically narrow than the elements of the other external sources. Also, mTwconv was the best with topic category \textit{Food and Beverages [J02]}, which we believe can be due to conversations about nutrition on Twitter. It is interesting that, on the other hand, pTwconv had no topic category with absolute Purity difference 0.2 or higher, and that the topic categories in mTwconv with the greatest improvement (\textit{Food and Beverages [J02]} and \textit{Geographicals [Z]}) are not the same as in pTwconv (which are \textit{Natural Science Disciplines [H01]}, \textit{Social Sciences [I01]} and \textit{Named Groups [M]}). This low absolute difference and difference in topic categories suggests that pTwconv needs the support of bTwconv to make a good clustering, which might be related to the low number of edges in the network (e.g. pTwconv has about two edges per external source element, while pTwauthor has about twenty). The profile of \textit{Food and Beverages [J02]} and \textit{Geographicals [Z]} in mTwconv is slightly higher than bTwconv (Figure \ref{figure:Examples of Purity profiles of several topic categories.}), which suggests that mTwconv is very competitive. On the other hand, the profiles in pTwconv are substantially lower, which is unfortunate because this is one of the networks that had the greatest improvements in \textit{Natural Science Disciplines [H01]} and \textit{Social Sciences [I01]}.

\subsection{Document Authors}

The network pAuthor was the best for the topic category \textit{Geographicals [Z]}, although it did poorly for the other topic categories. We believe that it was the best due to document authors having stable interests over time about given geographical regions. Figure \ref{figure:Examples of Purity profiles of several topic categories.} shows that \textit{Geographicals [Z]} achieve a substantially higher profile in pAuthor than in bAuthor or mAuthor, making it very competitive. This is especially interesting given that, based on our prior work \cite{previouswork}, the topic category \textit{Geographicals [Z]} is the worst topic category for text similarity and citation networks by a substantial margin. Document authors are already used to create science maps, but unlike our paper, the clusters in those maps tend to be made up of authors instead of documents, and the edges represent how many documents the authors have written together \cite{kumar2015co}.

\subsection{Facebook users}

pFacebook did well with topic category \textit{Named Groups [M]}, especially its topics about medical personnel, and was the best with \textit{Health Occupations [H02]}, especially for its topics medical specialties and nursing. This suggests that the users of Facebook are very interested in sharing documents related to health advice, which makes sense because it has a lot of support groups for people that suffer certain diseases where they share advice. The profile of mFacebook in these topic categories was higher than pFacebook, and for \textit{Health Occupations [H02]} it was about half that of bFacebook (Figure \ref{figure:Examples of Purity profiles of several topic categories.}), so we believe mFacebook to be competitive for \textit{Health Occupations [H02]}. It is worth mentioning that the highest topic profiles of pFacebook within the topic categories \textit{Named Groups [M]} and \textit{Health Occupations [H02]} were much higher than in bFacebook. Therefore, if there was a topic category that was exclusively composed of topics medical personnel, specialties and nursing, then this topic category would certainly have a much higher profile for pFacebook and mFacebook than for bFacebook. This shows that the topic categories that we use in the current paper might be insufficient to capture the benefit of the external sources.

\subsection{Policy documents}

pPolicy did well in the topic categories \textit{Named Groups [M]} and \textit{Geographicals [Z]}, and it was one of the few that had an improvement in \textit{Psychiatry and Psychology [F]}, although the improvement was small. We found that the theme that unifies the topics with a high Purity profile in each topic category were: In \textit{Psychiatry and Psychology [F]}, topics relevant to the government (e.g. combat disorders) or society (e.g. social phobia). In \textit{Named Groups [M]}, medical professions and vulnerable groups (e.g. undocumented immigrants, persons with mental disabilities, minors). In \textit{Geographicals [Z]}, American states and global south countries. The high Purity profiles in the former two topic categories are about government and social issues, which makes sense given that this is policy documents, while the topics from the latter one are relevant to the American government, which makes sense because the database has a better coverage of policy documents from the Anglo–Saxon world \cite{pinheiro2021large}. The profile of  \textit{Named Groups [M]} and \textit{Psychiatry and Psychology [F]} in pPolicy is substantially lower than in bPolicy, but the opposite is true in the profile of \textit{Geographicals [Z]} (Figure \ref{figure:Examples of Purity profiles of several topic categories.}). However, the Purity of the profile of \textit{Geographicals [Z]} is still extremely low, making it not very useful for science map users. Interestingly, the profile for mPolicy is lower than for either pPolicy and bPolicy, which is uncommon, suggesting that in this topic category the BERT and Pure networks do not complement each other to create better clusters.

\subsection{Patent families}

mPatents did well with the topic categories \textit{Chemicals and Drugs [D]}, especially its topics about biochemical elements, and \textit{Technology, Industry, and Agriculture [J01]}, especially its topics about chemical components. This suggests that this network does well for topics about Biotechnology, which is likely related to the inventions proposed in the patents. For the profile of \textit{Chemicals and Drugs [D]} and \textit{Technology, Industry, and Agriculture [J01]}  (Figure \ref{figure:Examples of Purity profiles of several topic categories.}), mPatents is about half that of bPatents, which we believe is enough for mPatents to be competitive. On the other hand, pPatents did poorly in absolute Purity difference.

\subsection{Twitter Authors}

The network pTwauthor was one of the best for \textit{Social Sciences [I01]} and \textit{Health Care [N]}, especially for topics about nursing. The reason for this topic to have a high clustering effectiveness is likely to be related to being one of the most shared topics in social media \cite{fang2020extensive}, but we also believe that it is due to a substantial number of Twitter users sharing documents exclusively related to nursing. Unfortunately, neither pTwauthor nor mTwauthor had topic categories with absolute Purity difference higher than 0.2, and the pTwauthor profile in these categories was substantially lower than in bTwauthor (Figure \ref{figure:Examples of Purity profiles of several topic categories.}), suggesting that pTwauthor is not competitive. Considering how well mTwconv did, this suggests that Twitter networks are more helpful for science maps when the networks are based on conversations instead of users, even as the second are more commonly used \cite{costas2021heterogeneous}. This could be because Twitter users tweet about multiple topics, while conversations are likely to be more topically focused. pTwauthor also did much worse than pFacebook, which is the other network with social media users as nodes, and we believe this can be due to Twitter having a high number of bot accounts that share academic documents automatically, at least compared with Facebook.

\subsection{Twitter networks versus the other networks}

We noticed that the Pure Twitter networks (pTwconv and pTwauthor) provide a very different perspective from the other sources. On one hand, if we ignore \textit{Organisms [B]}, these are the networks with the highest number of topic categories with a high relative Purity difference. On the other hand, these are the networks that achieved the highest improvement for topic category \textit{Natural Science Disciplines [H01]}, which is a category that science map users expect to see in science maps and that the traditional sources for science maps are not good at showing \cite{previouswork}. We believe that the perspective of Twitter is due to a dichotomy on how science is organized: On one hand, we have the social construction of how people think science should be organized, represented by Twitter and Facebook, and on the other hand we have organization that emerges from practical uses of science, represented by all the other sources. 

\section{Discussion}

In the current Section we will discuss the high level ideas, strengths and weaknesses of our work.

\subsection{High level ideas}

One of our most important results is that the external sources tend to cluster some topic categories better than others, and that these topic categories are different between sources. This suggests that external sources provide complementary perspectives on how to group documents together, and that these perspectives also have a meaning. These different perspectives are not only useful to create science maps, like in this paper, but they could potentially be applied in other areas to reveal how society perceives and engages with science. For example, that the Twitter perspective is very different from the other networks, or that Facebook users share health science, or that document authors are conservative about the geographical area they publish about. Also, even as the external sources tend to not outperform BERT in most topic categories, this was not the goal of the paper, and it is possible that an alternative method for constructing science maps could reach this goal.

The method that we have developed in this work can also be applied in other contexts. While our experiments focused on the biomedical field, the evaluation framework is not specific to it. In principle, the clustering effectiveness of any set of documents can be assessed using our method, regardless of discipline or the specific criteria used to define the set. We focused on biomedicine because MeSH terms provide an exceptionally well–maintained and complete classification system. 

Applying the method to other disciplines presents challenges that lie not in the method itself but in data availability. First, accurate document labeling is essential, since errors in labeling directly affect the reliability of Purity and NSC, and may also cause relevant documents to be excluded from the science map altogether, distorting cluster formation. Second, many classification systems outside biomedicine tend to lack the hierarchical structure of the MeSH tree, or have a highly unbalanced distribution of topics across categories, complicating the construction of topic categories. These are not insurmountable challenges, but they would require innovations in data curation and evaluation methodology.

The proposed evaluation framework is not limited to comparing science maps derived from different data sources. It can also be used to evaluate the extent to which a cluster that emerges in one science map also emerges in another science map. This is useful for assessing the stability of a cluster across different clustering algorithms or across time snapshots as science maps are updated with newly published documents. This can be done by treating the set of documents belonging to the cluster of interest as a “topic” and constructing its Purity profile.

\subsection{Strengths and weaknesses}

One of the motivations of our research is that clustering cannot be perfect because documents with multiple topics are assigned to a single cluster. However, this limitation can be addressed through different science mapping approaches. One approach is soft clustering, in which documents may belong to multiple groups simultaneously. This approach represents the topics of documents better, but it is not used in science map visualizations due to the difficulty of producing readable visualizations. Another approach is irreductionist mapping, which visualizes all documents to allow a more open interpretation of their topical relationships \cite{218bb68c3d354457bdd2f977f76c107f}, making clear which documents are close to the boundaries between topics. We do not study these approaches here because our analysis focuses on the most widely used paradigm in science mapping, namely hard clustering.

A strength of our research is the clustering effectiveness evaluation method, which is a substantial improvement over the clustering effectiveness evaluation method we used in our prior work \cite{previouswork} because our new approach is much easier to interpret. We used to have two metrics to evaluate effectiveness, Purity and the inverse clustering count, while now we only have Purity. We also used to only be able to compare clustering effectiveness between clustering solutions with the same documents and similar cluster sizes, while now we can compare the clustering solutions of several Resolution values across networks with different documents. In the prior work we also did not have Purity profiles, which provide a very intuitive description of the quality of the topic clusters that a user would experience in a science map. On the other hand, our evaluation method misses some of the nuance of our last work. For example, we did not evaluate if some sources are better than others at different cluster sizes (our prior work and Xie and Waltman \cite{xie2023comparison} found that citations are better than text for smaller clusters).

A limitation of our work is that we performed our experiments on clustering solutions that are less sophisticated than science maps used by researchers. For example, some science map methodologies have a minimum size for clusters, and clusters smaller than this size are merged with other clusters \cite{waltman2012new}. We did not do this, and as a consequence, when the nodes of a cluster are all equally connected by a few hub nodes in the network, reducing the size of the cluster by increasing the Resolution will turn random nodes of this cluster into singletons. This is a problem because, if this node is a topic document, then Purity would decrease at higher NSC, creating very confusing results for some topics that do not reflect the cluster effectiveness that would be observed in a science map. We observed this situation mostly in the Twitter users source, where some documents were shared by only one or two users. We did not attempt to prevent this situation because doing so would increase the complexity of our experimental design.

Another limitation of our research is the possibility of shared signals between the text–based and citation–based similarity networks. Our text embeddings were generated using the allenai–specter model, which is based on the SPECTER BERT–like model. SPECTER was originally trained on a corpus that included citation information, with the objective that documents connected by citation links would have more similar embeddings. As a result, the text similarity network may implicitly encode some citation–related structure. This remains a plausible interpretation rather than a demonstrated effect. We consider this a pragmatic compromise, as SPECTER was designed specifically to represent the semantics of academic documents.

Another limitation is that our Mixed networks combine a non–bipartite network (the BERT networks, non–bipartite because the links go from document to document) with a bipartite network (the Pure networks, bipartite because the links go from document to external source element). There are studies that use either of these types of networks for creating science maps, but there are no studies about combining them, which could have unintended effects in the map. The closest there is in the literature is the extended citation networks, where there are links from document to document and from document to non–core document, but not from non–core document to non–core document. Also, bipartite networks are not very common in science mapping, and it is more common to, instead of having the unit of co–occurrence in the network (in our case, the external source element), to represent the co–occurrence in the edge weight \cite{small1973co}. The method we used to combine the networks into the Mixed network is also relatively straightforward, and the only modification that we make is that the sum of edges weights in both networks must be the same. We can imagine alternative modifications, for example making all the edges that came out from a node to add up to the same value. We did not explore these alternatives to not further complicate our analyses, but future research could explore how to create better Mixed networks for a given external source.

Another limitation of this study is that some of the data sources we rely on may not be readily available to all researchers using science maps. For example, access to the Twitter API was previously free for academic research but has since become restricted to paid tiers. Such shifts in data accessibility are well documented in altmetrics research \cite{Haustein2016GrandChallengesAltmetrics}. More generally, the availability of data platforms depends on the policies of the organizations that control them, which may change over time. This issue is not limited to social media data, but applies to many types of external data sources. For this reason, future research should consider incorporating open data sources for the academic publications, such as OpenCitations \cite{opencitations_article}, OpenAlex \cite{priem2022openalex} and Crossref Event Data \cite{crossrefeventdata}. Such work could also evaluate whether the choice of data source for academic publications affects the topical biases of a science map. Despite this limitation, we argue that our results remain relevant. New data sources may become available in the future, and the framework presented in this work can be applied to evaluate their potential contribution to science maps, including identifying which topic categories are most likely to benefit from their integration.

\section{Conclusions}

The topical bias of science maps limits their usefulness for topical analyses. In the current paper we have explored different data sources for creating academic documents networks that represent different document relations, with the purpose of finding sources that can change the topical bias of a science map. Our method of analysis was comparing the clustering effectiveness of different MeSH topic categories within a network and between networks, using a methodology that we refined from our prior work. We explored traditional science maps data sources (text similarity and citation links) and non–traditional data sources based on the co–occurrence of academic documents on another element (policy document, patent families, Facebook users, Twitter conversations, Twitter users, and document authors), which we referred to as external sources. Our comparisons were between networks that use either text similarity, external sources, or a mix of both.

We found that different external sources can be used to favor the emergence of different topics, and the following combinations had a particularly strong effect: Health for Facebook users, biotechnology for patent families, government and social issues for policy documents, food for Twitter conversations, nursing for Twitter users, and most strongly geographical entities for document authors. We also found that Twitter conversations work particularly well when combined with text similarity, even as they perform poorly in their absence. Also, the favored topic categories are not affected by changing the percentage of the topic documents used in the evaluation, as shown by the similarity between the different Coverage values. Finally, the best topic categories in the Twitter networks were very different from the other networks, which means that Twitter (and potentially other similar social media platforms, like the new BlueSky or Mastodon) might provide different perspectives for the study of the organization of scientific knowledge, getting us closer to latent representations of how society perceives and interacts with science.

Our text similarity networks were constructed using Sentence–BERT embeddings generated with the allenai–specter model, which to our knowledge has not previously been used to create science maps. We found that this metric generally performed better than similarity measures used in prior work (such as BM25), with the exception of topics related to organisms. Moreover, combining Sentence–BERT similarity with citation networks produced the best overall clustering effectiveness. These findings suggest that text similarity–based science maps may benefit from using Sentence–BERT embeddings such as allenai–specter, particularly when integrated with citation–based signals.

Our results show that external sources of academic document networks can be used to control topic bias, which opens up the possibility of creating science maps tailored for different needs. The most direct way of applying our discoveries is to create science maps biased toward different topics using these external sources. However, with the exception of document authors and their high clustering effectiveness for geographical entities, most external sources need to be used in combination with text similarity sources to achieve a high clustering effectiveness relative to traditional sources, and it is still an open question which is the best method for combining them into a single network. The clusters of external sources could also be used beyond science maps, for example to identify potential misuse of scientific publications (e.g. in misinformation strategies), or to identify societal connections or sensitivities that are not reflected in the academic world (e.g. connecting papers of diets and health concerns).

Finally, while this study focused on biomedicine due to the availability of MeSH annotations, the proposed evaluation framework is not domain–specific and could be applied to other disciplines given appropriate document labeling.

\section*{Acknowledgments}

We thank MetaROR \cite{MetaROR} for providing an open peer review platform and for identifying suitable reviewers, as well as Daniel Hook, Silvio Peroni and Verena Weimer, who freely gave their time and effort to provide constructive and insightful feedback \cite{MetaROR_EA_34_1}.

\section*{Supplementary material} \label{Supplementary material}

The data and the code used to create the results is available at a Zenodo repository \cite{bascur_2024_14170722}. The original data cannot be shared because the Web of Science document identifiers are proprietary. To enable data sharing, we replaced the Web of Science identifiers with internally generated Paper IDs. These Paper IDs are used consistently across all shared datasets. Detailed usage instructions and additional clarifications are provided in the repository.

\bibliographystyle{aomplain}
\bibliography{mybib}

\end{document}